\documentclass[lettersize,journal]{IEEEtran}
\usepackage{amsmath,amsfonts}
\usepackage{algorithmic}
\usepackage{array}
\usepackage[caption=false,font=normalsize,labelfont=sf,textfont=sf]{subfig}
\usepackage{textcomp}
\usepackage{stfloats}
\usepackage{url}
\usepackage{verbatim}
\usepackage{graphicx}

\usepackage{etoolbox}

\makeatletter
\patchcmd{\@maketitle}
  {\centering}
  {\centering
   {\normalfont\scriptsize
    This work has been submitted to the IEEE for possible publication. Copyright may be transferred without notice, after which this version may no longer be accessible.\par}
   \vspace{1em}}
  {}{}
\makeatother

\def\BibTeX{{\rm B\kern-.05em{\sc i\kern-.025em b}\kern-.08em
    T\kern-.1667em\lower.7ex\hbox{E}\kern-.125emX}}
\usepackage{balance}
\begin{document}
\title{Pulling Illusion in Individuals with Neurological Disorders}
\author{Takeshi~Tanabe,
        Satoshi~Yamamoto,
        Toru~Yamada,
        Daisuke~Ishii,
        and~Yutaka~Kohno
\thanks{This work involved human subjects or animals in its research. Approval of all ethical and experimental procedures and protocols was granted by the Ethics Committee of the Ibaraki Prefectural University of Health Sciences under Approval No. 1107.}
\thanks{This work was supported by the JSPS KAKENHI Grant No. 26K23862, and JST, PRESTO Grant No. JPMJPR24I7.}
\thanks{T. Tanabe is with Research Institute on Human and Societal Augmentation, National Institute of Advanced Industrial Science and Technology (AIST), 6-2-3 Kashiwanoha, Kashiwa, Chiba 277-0882, Japan (e-mail: t-tanabe@aist.go.jp).}
\thanks{S. Yamamoto and Y. Kohno are with Ibaraki Prefectural University of Health Sciences, 4669-2 Ami, Ami-cho, Inashiki-gun, Ibaraki 300-0394, Japan (e-mail: yamamotos.jpn@gmail.com; kohno@ipu.ac.jp).}
\thanks{T. Yamada is with Human Informatics and Interaction Research Institute, National Institute of Advanced Industrial Science and Technology (AIST), 1-1-1 Umezono, Tsukuba, Ibaraki 305-8568, Japan (e-mail: toru.yamada@aist.go.jp).}
\thanks{D. Ishii is with Graduate School of Biomedical and Health Sciences, Hiroshima University, 1-2-3 Kasumi, Minami-ku, Hiroshima City, Hiroshima, 734-8553, Japan (e-mail: ishiid@hiroshima-u.ac.jp).
}}

\maketitle

\begin{abstract}
The pulling illusion induced by asymmetric vibration stimuli has attracted attention for its potential applications in rehabilitation and sensory assessment.
However, the underlying mechanism of the pulling illusion remains unclear.
This study addressed the central question of whether peripheral vibrotactile sensitivity alone is sufficient for the illusion to emerge or whether processing beyond basic vibration detection is also required.
Neurological disorders can involve impairments at different levels of the nervous system, providing an opportunity to examine this question.
Accordingly, we evaluated directional discrimination performance for the pulling illusion and fingertip vibration detection thresholds in 25 participants with diverse neurological disorders affecting different levels of the nervous system, from peripheral to central.
Clustering analysis identified contrasting profiles, with high directional discrimination performance despite elevated vibration detection thresholds and chance-level performance despite relatively low-to-intermediate thresholds.
In the generalized linear mixed model, motor-related signs, including hemiplegia and tremor, showed a robust negative association with directional discrimination performance, whereas vibration detection threshold was not robustly associated with performance.
Furthermore, in participants with hemiplegia, directional discrimination performance was around chance level on the affected side and close to 100\% on the unaffected side, despite stimulus amplitudes well above the measured vibration detection thresholds on both sides.
Collectively, these findings suggest that the pulling illusion depends on perceptual processing beyond basic vibration detection, through which asymmetric vibration is experienced as directional pulling.
\end{abstract}


\section{Introduction}
Haptic illusions are increasingly recognized for their potential in neurorehabilitation. 
Well-known examples include kinesthetic illusions induced by vibratory stimulation of tendons \cite{KI_reha_1, KI_reha_2} and the rubber hand illusion \cite{RHI_reha_1, RHI_reha_2}.
Among various haptic illusions, the pulling illusion \cite{amemiya2, rekimoto, tan_toh, 2ch, cl_chi, yeya, dong, donkov, sabnis} has attracted attention for its unique perceptual features. 
This illusion is typically induced by asymmetric waveforms that combine brief, rapid acceleration in one direction with longer, slower acceleration in the opposite direction.
This asymmetry produces the compelling illusion that the hand is being pulled in a specific direction \cite{amemiya2}.

The pulling illusion has recently been explored for applications in rehabilitation \cite{PI_reha, PI_gait, akashi2025} and sensory assessment \cite{fukazawa2025}.
Supporting its potential in these contexts, recent studies have demonstrated that it can interact with other haptic illusions used in rehabilitation, such as kinesthetic illusions \cite{noto2024} and rubber hand illusions \cite{akashi2025}, and can also unconsciously influence ballistic movement trajectories \cite{tan_tnsre2024}. 
These findings open up novel possibilities for motor training and rehabilitation design.
However, the mechanism through which this illusion emerges remains unknown, which is a major challenge.
Without understanding the underlying mechanism, its clinical application remains fundamentally limited. 

The conventional hypothesis regarding the mechanism of this illusion has focused on the idea that directionally biased deformation of the fingertip skin produced by asymmetric vibration is detected as a pulling sensation \cite{ame_eh,cl_hs}.
According to this account, directional discrimination performance for the illusion would be expected to depend on peripheral vibrotactile sensitivity involved in detecting vibration-induced skin deformation at the fingertip.
However, our previous studies showed that vibration detection threshold was not consistently associated with directional discrimination performance for the illusion \cite{tan_toh2025, tan_toh2026}.
Moreover, a neurophysiological study showed that delayed activity in the contralateral parietal cortex, approximately 300 ms after stimulation, was associated with the pulling illusion \cite{havas}.
These findings suggest that, in addition to basic vibration detection at the fingertip as proposed by the conventional hypothesis, the pulling illusion may also depend on subsequent perceptual processing.

To explore the possible contribution of processing beyond basic vibration detection to the pulling illusion, this study focused on individuals with neurological disorders, in whom impairments can involve different levels of the nervous system, ranging from peripheral sensory pathways to central nervous system processing.
Evaluating the illusion across such heterogeneous impairments provides an opportunity to examine whether directional discrimination performance can be preserved despite impaired vibration detection or weakened despite relatively preserved vibration detection.
In this study, we evaluated directional discrimination performance for the pulling illusion and examined its association with fingertip vibration detection thresholds across a heterogeneous population of individuals with neurological disorders.
Through this approach, we addressed whether peripheral vibrotactile sensitivity alone is sufficient for the pulling illusion to emerge or whether functions associated with neurological disorders are also involved in its emergence.
To our knowledge, this is the first study to evaluate directional discrimination performance for the pulling illusion in participants with neurological disorders, allowing us to examine the illusion across neurological variability not available in studies of healthy participants alone.

\begin{table*}[t!]
\caption{Participant list.}
\centering
\label{tab:p}
\scalebox{1}{ 
\begin{tabular}{l|l|l|l|l|l} \hline
ID& Age& Gender& Disease name& Lesion location& Motor-related signs (hemiplegia or tremor) \\ \hline
P1& 58& F& Carpal tunnel syndrome& &  \\
P2& 63& F& Carpal tunnel syndrome& &  \\
P3& 78& F& Carpal tunnel syndrome& &  \\
P4& 83& F& Carpal tunnel syndrome& &  \\
P5& 53& F& Carpal tunnel syndrome& &  \\
P6& 50& F& Carpal tunnel syndrome& &  \\
P7& 58& F& Carpal tunnel syndrome& &  \\
P8& 67& M& Peripheral neuropathy& &  \\
P9& 61& M& Subarachnoid hemorrhage& Right medulla oblongata& \\
P10& 77& F& Intracranial hemorrhage& Left capsular & \\
P11& 63& M& Intracranial hemorrhage& Right thalamus & With left hemiplegia\\
P12& 59& M& Intracranial hemorrhage& Left capsular & With right hemiplegia\\
P13& 55& M& Intracranial hemorrhage& Left capsular & With right hemiplegia\\
P14& 67& F& Intracranial hemorrhage& Right thalamus & With left hemiplegia\\
P15& 68& M& Cerebral infarction& Frontal, parietal, and temporal lobes & With left hemiplegia\\
P16& 55& F& Cerebral infarction& Frontal and parietal lobes& With right hemiplegia\\
P17& 71& F& Multiple system atrophy& & \\
P18& 61& F& Spinocerebellar degeneration& & \\
P19& 77& M& Spinocerebellar degeneration& & \\
P20& 79& F& Parkinson's disease&  & With tremor \\
P21& 68& F& Parkinson's disease&  & \\
P22& 74& F& Parkinson's disease&  & \\
P23& 42& M& Parkinson's disease&  & With tremor \\
P24& 75& F& Parkinson's disease&  & \\
P25& 59& F& Muscular dystrophy&  & \\
\hline
\end{tabular}
}
\end{table*}

\section{Methods}
\subsection{Participants}
A total of 25 participants with neurological disorders, aged 42--83 years (mean: 64.8 $\pm$ 10.3 years; 17 females), were enrolled in this study (TABLE~\ref{tab:p}). 
Participants were recruited from the Ibaraki Prefectural University of Health Sciences Hospital in Japan. 
The cohort was intentionally designed to include individuals with a wide range of neurological impairments spanning from the peripheral to the central nervous system, ensuring variability in functionally relevant impairments. 
The breakdown of participants by condition was as follows: seven with carpal tunnel syndrome (CTS), one with peripheral neuropathy, one with subarachnoid hemorrhage (SAH), five with intracranial hemorrhage (ICH), two with cerebral infarction (CI), one with multiple system atrophy (MSA), two with spinocerebellar degeneration (SCD), five with Parkinson's disease (PD), and one with muscular dystrophy.
Among the manifestations of neurological disorders observed in this cohort, we focused on motor-related signs that could be consistently identified at the side level across participants.
Motor-related signs were defined as motor abnormalities relevant to the tested side, specifically hemiplegia or tremor in the present cohort, and were included in the statistical model.
Exclusion criteria included severe motor or sensory impairment that prevented task execution, as well as conditions that interfered with informed consent. 
This study was approved by the Ethics Committee of the Ibaraki Prefectural University of Health Sciences (approval number: 1107), and the experimental procedures were conducted in accordance with the Declaration of Helsinki. 
Informed consent was obtained from all participants.
All participants completed the experimental procedures.
Two participants were subsequently excluded from the statistical analyses according to the response-pattern exclusion criterion described in Sec.~II-E.

\subsection{Study Design}
The study was designed to evaluate the relationship between directional discrimination performance for the illusion and functionally relevant factors across participants.
Conceptually, the study was based on the following regression framework:
\begin{equation}
\begin{aligned}
Y(i,j) \sim \beta_0
&+ \beta_{VT} X_{VT}(i,j) \\
&+ \beta_{MS} X_{MS}(i,j) \\
&+ \beta_{A} X_{A}(i) \\
&+ \beta_{G} X_{G}(i) \\
&+ u(i) ,
\label{eq:conceptual_model}
\end{aligned}
\end{equation}
where $i$ denotes the participant, $j$ denotes the tested side, and $u_i$ represents participant-specific variability.
The coefficients $\beta_0$, $\beta_{VT}$, $\beta_{MS}$, $\beta_A$, and $\beta_G$ represent the intercept and the effects of vibration detection threshold, motor-related signs, age, and gender, respectively.
The outcome variable, $Y(i,j)$, represents directional discrimination performance for the illusion.
$X_{VT}(i,j)$ represents vibration detection threshold as an index of peripheral vibrotactile sensitivity, whereas $X_{MS}(i,j)$ represents the presence of motor-related signs on tested side $j$ of participant $i$, specifically hemiplegia or tremor in the present cohort.
$X_A(i)$ and $X_G(i)$ represent participant-level age and gender, respectively.
Accordingly, the pulling illusion was evaluated using a directional discrimination task, and basic sensory function was quantified by measuring vibration detection thresholds at the fingertips.
Both measures were obtained bilaterally, allowing side-specific analysis.
The tested side was treated as the unit of analysis, and each participant therefore contributed up to two observations corresponding to the left and right sides.
Motor-related signs were coded separately for each tested side; for participants with unilateral motor symptoms, the affected side was coded as having motor-related signs.
This design allowed us to examine whether directional discrimination performance for the illusion was primarily associated with vibration detection threshold or whether motor-related signs provided additional information.

\subsection{Experimental setup}
\begin{figure}[!t]
\begin{center}
\includegraphics[width=8cm,keepaspectratio,clip]{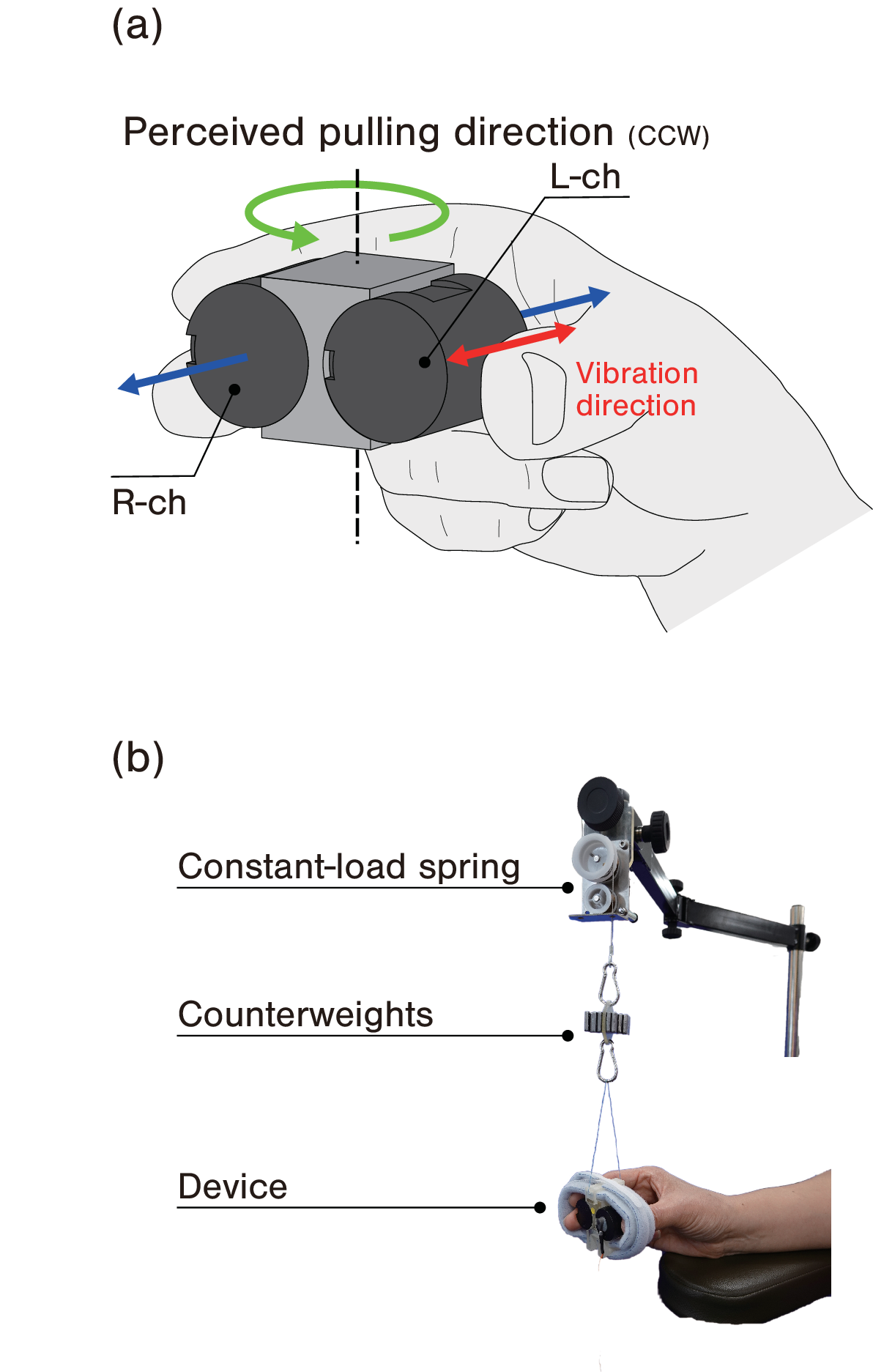}
\caption{Experimental setup. (a) Configuration diagram of vibratory device (size: 53(w)$\times$28(h)$\times$27(d) mm, weight: 67.4 g). (b) Suspended vibratory device.}
\label{fig:dev}
\end{center}
\end{figure}

The evaluation of directional discrimination performance for the illusion and vibration thresholds was conducted using an identical experimental setup. 
Vibratory stimuli were generated using voice coil-type actuators (639897, Foster Electric Co., Ltd.), as illustrated in Fig.~\ref{fig:dev} (a).
To induce the pulling illusion in rotational directions, two actuators (L-ch and R-ch) were mounted in parallel on a custom fixture fabricated with a 3D printer \cite{2ch,cl_chi}.
The control signals were generated in MATLAB R2020b (MathWorks Inc.) and converted into audio signals using Psychtoolbox, a software library designed for psychophysical experiments. 
These signals were then output through an audio interface (PAA-U4P, MIYOSHI) and amplified by a Class-D amplifier (PAM8403DR, Diodes Inc.). 
To measure the vibrations, an accelerometer (NP-2110, Ono Sokki Co., Ltd.) was attached to the actuator's surface using adhesive and thumb contact. It was connected to an amplifier (CH-1200A, Ono Sokki Co., Ltd.). 
A multifunctional data acquisition (DAQ) device (USB-6003, National Instruments Co.) recorded the measured accelerations at a sampling rate of 20 kHz. 
For rapid deployment in clinical settings, all components were integrated into a portable plastic enclosure. 
The system can be operated simply by connecting the enclosure to a power outlet and linking it to a laptop via a USB cable.

The participants were seated comfortably in a chair and held the vibratory device with their thumb, index and middle fingers. 
Since many participants with motor impairments were likely unable to grasp the device stably on their own, their fingers were secured with a soft band made of polyurethane (8197M, Medical Project Co., Ltd.) as shown in Fig.~\ref{fig:dev} (b). 
Additionally, to minimize the load on the participants, the device was suspended using a constant-load spring (4550, SAMINI Co., Ltd.), with its tension adjusted to 14.0 gf by counterweights. 
This tension was determined empirically to make the device stable and easy to hold while minimizing its effect on the illusion. 
These grasping methods were performed on all participants, regardless of disease, unaffected or affected side. 
The arm grasping the device was placed on the armrest in a comfortable position.

\subsection{Procedure}
The experiment began with the measurement of vibration thresholds, followed by the directional discrimination of the illusion. 
Both tests were conducted bilaterally for all participants, regardless of their condition.
For participants with hemiplegia, measurements were first taken on the unaffected side, followed by the affected side.
For all other participants, the order of testing was randomized.

\subsubsection{Vibration thresholds}
The absolute threshold for sinusoidal vibration at 75 Hz, corresponding to the fundamental frequency of the asymmetric vibration stimuli used to induce the illusion \cite{tan_tmech}, was measured. 
Since the experiment was conducted following an outpatient visit, the method of limits was used to measure vibration thresholds efficiently. 
To establish an appropriate stimulus range, an approximate vibration threshold was first estimated using the method of adjustment, with the maximum stimulus intensity set at twice this threshold. 
Vibration stimuli lasting 1 second were presented in steps of 10\% relative to the maximum intensity, increasing from the no-stimulus condition and decreasing from the maximum intensity, to determine the vibration detection threshold.
Participants verbally indicated whether they detected the vibration.
Two ascending and two descending trials were conducted, with stimulus intensity adjusted until a change in the participant's response was detected in each trial. 
The time-series vibration data, recorded using an accelerometer, were analyzed via fast Fourier transform (FFT).
The vibration threshold was determined as the average of four trials.

\subsubsection{Directional discrimination of the pulling illusion}
\begin{figure}[!t]
\begin{center}
\includegraphics[width=8.5cm,keepaspectratio,clip]{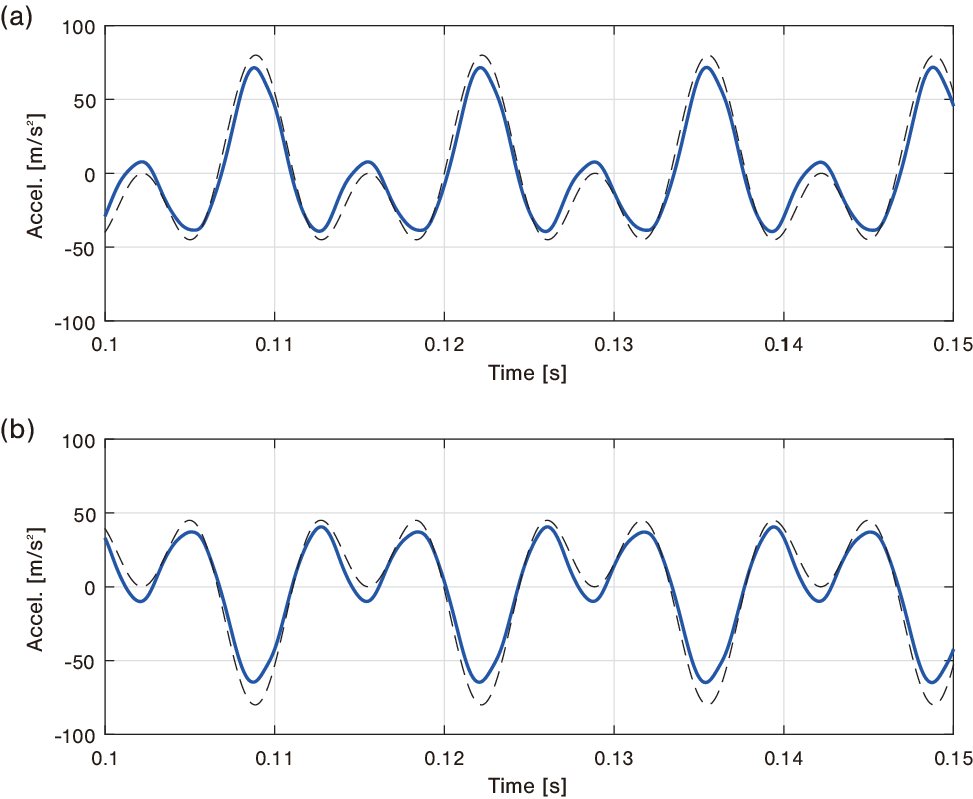}
\caption{Examples of asymmetric vibration stimulus time-series data. Solid lines represent measured values, and broken lines represent the target. (a) Stimulus designed to induce the pulling illusion in the CW direction. (b) Stimulus designed to induce the pulling illusion in the CCW direction.}
\label{fig:wave}
\end{center}
\end{figure}

Directional discrimination performance for the pulling illusion was defined as the correct response rate in the directional discrimination task.
As in a previous study examining the effects of aging on illusions \cite{tan_toh2025}, the pulling illusion was induced in the clockwise (CW) and counterclockwise (CCW) directions by generating opposing translational illusory forces between the L-ch and R-ch actuators, arranged in parallel \cite{2ch,cl_chi}, as shown in Fig.~\ref{fig:dev} (a).
A representative example of the asymmetric vibration stimulus is shown in Fig.~\ref{fig:wave}.
This waveform, previously used to evaluate the pulling illusion \cite{tan_toh2025}, consisted of 75 Hz and 150 Hz sine waves with an acceleration amplitude of 40 $\mathrm{m/s^2}$, combined. 
By reversing the sign of this waveform, the pulling illusion was induced in either the CW or CCW direction.

Participants grasped the device in the same manner as during the vibration threshold  measurement and were randomly presented with asymmetric vibration stimuli designed to induce the pulling illusion in either the CW or CCW direction.
A two-alternative forced-choice (2AFC) paradigm was employed to assess directional discrimination performance, with a chance level of 50\%.
Participants verbally reported the perceived direction. 
Each asymmetric vibration stimulus was presented for one second, followed by a two-second interval before the next stimulus was delivered, after the participant's response. 
A total of 40 trials were conducted, with 20 trials for each direction.

\subsection{Data Analysis}
\begin{figure*}[!t]
\begin{center}
\includegraphics[width=17.5cm,keepaspectratio,clip]{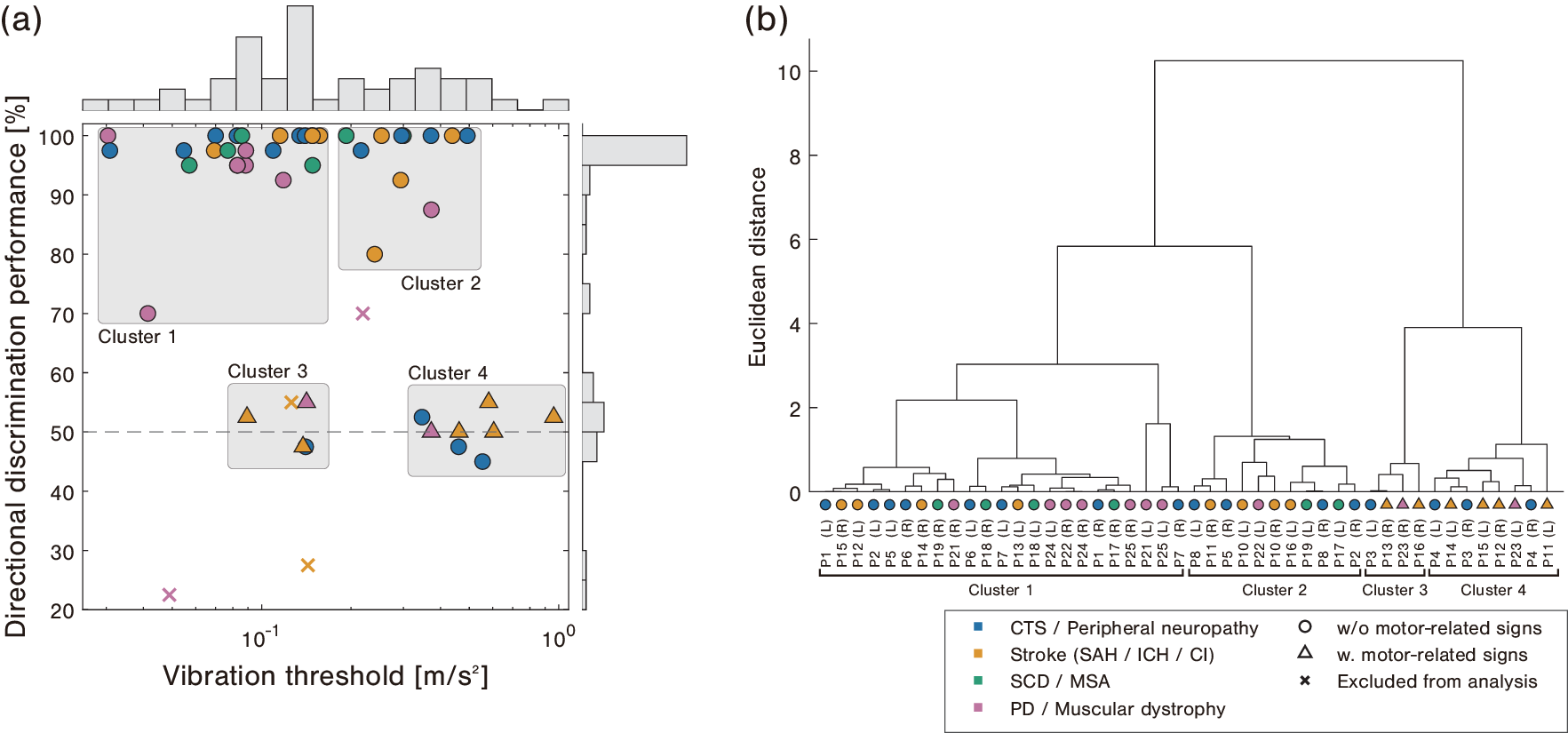}
\caption{Relationship between vibration threshold and directional discrimination performance for the pulling illusion. Marker color indicates disease category, whereas marker shape distinguishes observations with and without motor-related signs. Crosses indicate observations from the two participants shown descriptively but excluded from the clustering and regression analyses because of markedly below-chance performance on one tested side.
(a) Scatter plot showing all side-level observations, with marginal histograms of vibration threshold and directional discrimination performance. The dashed line denotes chance level (50\%).
(b) Hierarchical clustering dendrogram based on the observations included in the analysis.}
\label{fig:scatter}
\end{center}
\end{figure*}

Before the statistical analyses, participants were excluded if directional discrimination performance on either tested side was markedly below chance level.
This criterion was applied at the participant level because the present analyses interpreted directional discrimination performance along a continuum from chance-level responding to consistently correct discrimination.
Two participants met this criterion: P9 and P20, who showed correct response rates in the directional discrimination task of 27.5\% and 22.5\%, respectively, on one tested side.
Of these, P20 had motor-related signs.
Both sides from each of the two participants were therefore excluded from the analyses, resulting in four excluded observations.
These four excluded observations are shown descriptively in the scatter plot in Fig.~\ref{fig:scatter} (a).

To examine side-level response tendencies, hierarchical clustering analysis was performed using vibration threshold and directional discrimination performance.
The vibration threshold was log-transformed (base 10), and both variables were $z$-standardized before clustering. 
Pairwise distances were calculated using Euclidean distance, and clusters were generated using Ward's linkage method. 
Based on the dendrogram structure and the interpretability of the resulting response profiles, a four-cluster solution was used to summarize side-level tendencies in the combinations of vibration threshold and directional discrimination performance.

After characterizing these distributional profiles, we next evaluated whether directional discrimination performance was statistically associated with vibration threshold and motor-related signs.
Based on the conceptual regression framework described in Eq. (\ref{eq:conceptual_model}), the number of correct responses in the directional discrimination task for the illusion was analyzed using a binomial generalized linear mixed model (GLMM), with the total number of trials per condition specified as the denominator.
Fixed effects included vibration threshold, age, gender, and the presence of motor-related signs, and participant was included as a random intercept. 
Continuous predictors were $z$-standardized before model fitting to facilitate comparison of coefficient estimates.
All analyses were conducted in R (version 4.5.3) using the lme4 package, with the bobyqa optimizer.
To evaluate the robustness of the fixed effects, a participant-level stratified bootstrap analysis was performed. 
Participants were resampled with replacement while preserving the proportion of motor-related signs (7 of 23 participants exhibited motor-related signs), and both sides from each resampled participant were retained as a cluster. 
For each bootstrap sample, the GLMM was refitted, and percentile-based 95\% confidence intervals (CIs) were calculated from 2000 bootstrap iterations. 
Effects were considered robust when the CI did not include zero.

As a supplementary analysis, participants with hemiplegia (ID: P11--P16), for whom motor-related signs were unilateral, were analyzed separately to examine whether side-specific differences were observed within the same participants. 
Comparisons were made between the affected and unaffected sides for vibration threshold and directional discrimination performance.
Since normality was rejected for both variables (Shapiro--Wilk test), the Wilcoxon signed-rank test was used.

\section{Results}
After participant-level exclusion, the clustering and GLMM included 46 side-level observations from 23 participants.
Fig.~\ref{fig:scatter} shows the relationship between vibration threshold and directional discrimination performance for the pulling illusion, with hierarchical clustering results overlaid.
Directional discrimination performance showed a bimodal-like distribution, with many observations located near 100\% and another group located around the chance level (50\%).
This pattern was also reflected in the marginal histogram of directional discrimination performance.

Hierarchical clustering summarized the side-level response tendencies into four clusters.
Cluster 1 showed high directional discrimination performance with relatively low vibration thresholds, whereas Cluster 4 showed chance-level directional discrimination performance with relatively elevated thresholds.
In contrast, Cluster 2 showed high directional discrimination performance despite elevated vibration thresholds, whereas Cluster 3 showed chance-level directional discrimination performance despite relatively low-to-intermediate vibration thresholds.
Observations with motor-related signs, indicated by triangular markers in Fig.~\ref{fig:scatter}, were frequently included in the chance-level clusters, particularly Clusters 3 and 4.
In addition, some observations from participants with CTS were included in the chance-level clusters.

\begin{figure}[!t]
\begin{center}
\includegraphics[width=8cm,keepaspectratio,clip]{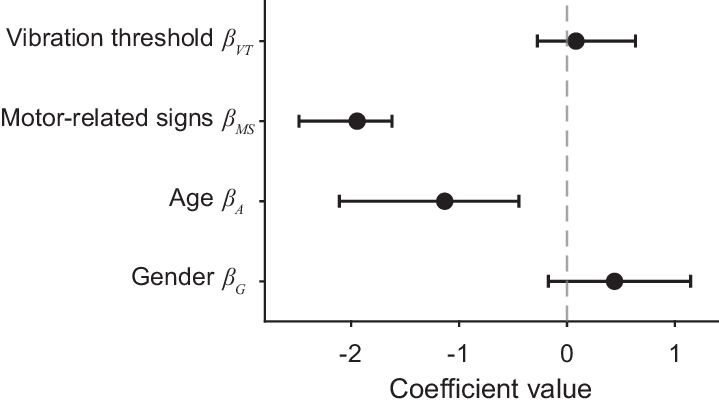}
\caption{Coefficients estimated using the generalized linear mixed model. Points indicate coefficient estimates on the log-odds scale, and horizontal error bars indicate percentile-based 95\% CIs. Continuous predictors were $z$-standardized before model fitting, whereas categorical predictors indicate contrasts relative to the reference category. The vertical dashed line indicates zero.}
\label{fig:estim}
\end{center}
\end{figure}

To quantitatively evaluate whether directional discrimination performance for the illusion was associated with vibration threshold or motor-related signs, a binomial GLMM was fitted based on the conceptual framework shown in Eq.~(\ref{eq:conceptual_model}).
The robustness of the fixed effects was then evaluated using bootstrap 95\% CIs (Fig.~\ref{fig:estim}).
Motor-related signs showed a robust negative association ($\beta_{MS} = -1.95$, 95\% CI [-2.48, -1.62]), with the CI not including zero.
In contrast, the coefficient for vibration threshold was close to zero ($\beta_{VT} = 0.08$, 95\% CI [-0.28, 0.64]), and its CI crossed zero, indicating that the direction of the association was not robust.
In addition, age showed a negative association, with a CI that did not include zero ($\beta_A = -1.13$, 95\% CI [-2.11, -0.45]), whereas the CI for gender included zero ($\beta_G = 0.44$, 95\% CI [-0.17, 1.15]).
In summary, motor-related signs showed a robust negative association with directional discrimination performance for the pulling illusion, whereas vibration threshold showed an estimate close to zero and no robust direction of association.

\begin{figure}[!t]
\begin{center}
\includegraphics[width=7cm,keepaspectratio,clip]{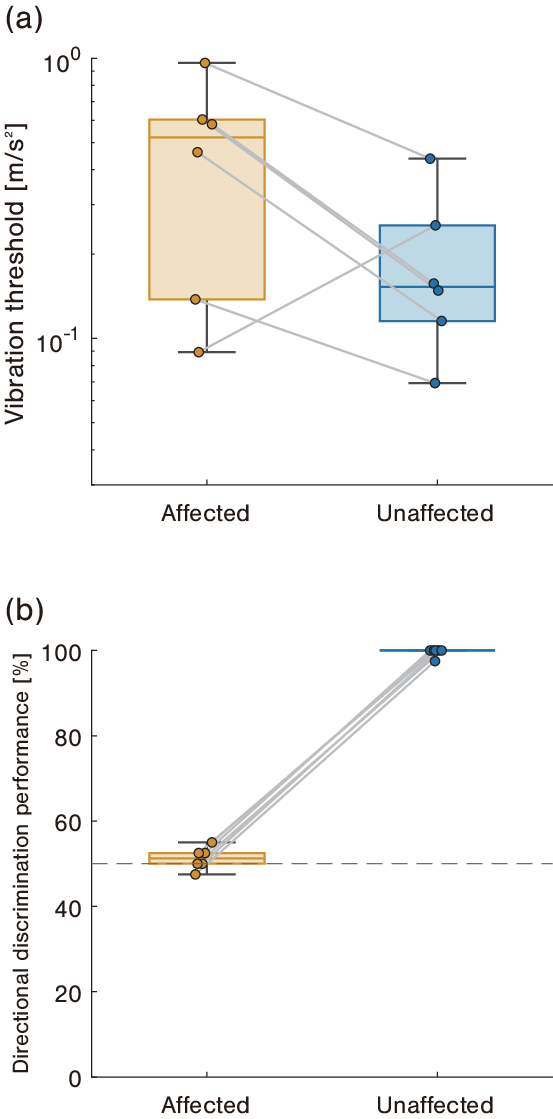}
\caption{Comparison between the affected and unaffected sides in participants with hemiplegia. (a) Vibration thresholds. (b) Directional discrimination performance for the pulling illusion.}
\label{fig:side_diff}
\end{center}
\end{figure}

Next, because the GLMM showed a robust negative association between motor-related signs and directional discrimination performance, we performed a supplementary comparison between the affected and unaffected sides in participants with hemiplegia.
By comparing the affected and unaffected sides, we examined whether motor-related signs reduced directional discrimination performance.
Fig.~\ref{fig:side_diff} shows paired comparisons of vibration thresholds and directional discrimination performance for the pulling illusion between the affected and unaffected sides in participants with hemiplegia.
Although a trend was observed suggesting that the vibration threshold was higher on the affected side than on the unaffected side, no significant difference was found between the two ($p = 0.09$). 
In contrast, a significant difference in directional discrimination performance was observed between the affected and unaffected sides ($p = 0.03$).
On the affected side, directional discrimination performance was consistently around chance level (50\%) across participants, whereas it was close to 100\% on the unaffected side.
Therefore, within participants with hemiplegia, directional discrimination performance for the illusion showed a clearer and more consistent side difference than vibration threshold.

\section{Discussion}
\subsection{Motor-related signs and directional discrimination performance for the pulling illusion}\label{sec:dis_hyp}
The central question of this study was whether peripheral vibrotactile sensitivity alone is sufficient for the pulling illusion to emerge or whether functions associated with neurological disorders are also involved in its emergence.
The clustering analysis provided evidence against the former possibility.
If peripheral vibrotactile sensitivity were the primary determinant of whether the pulling illusion emerges, the data should form two profiles: Cluster 1, characterized by a low threshold and high directional discrimination performance, and Cluster 4, characterized by a high threshold and chance-level directional discrimination performance.
However, the clustering analysis also identified Cluster 2 (high threshold and high directional discrimination performance) and Cluster 3 (low-to-intermediate threshold and chance-level directional discrimination performance).
The profiles of Clusters 2 and 3 indicate that peripheral vibrotactile sensitivity alone was insufficient to account for directional discrimination performance for the pulling illusion.

The results of the GLMM directly addressed this central question.
Vibration detection threshold showed no robust association with directional discrimination performance, whereas motor-related signs showed a robust negative association.
This contrast supports the possibility that functions associated with neurological disorders, as indicated by motor-related signs, are involved in the emergence of the pulling illusion beyond basic vibration detection.

This interpretation was further supported by the comparison between the affected and unaffected sides in participants with hemiplegia.
Although vibration thresholds did not differ significantly between sides, directional discrimination performance was consistently around chance level on the affected side and close to 100\% on the unaffected side.
Importantly, the acceleration amplitude of the asymmetric vibration stimuli used to induce the illusion was approximately 10--100 times higher than the measured vibration detection thresholds (see Figs.~\ref{fig:wave} and~\ref{fig:side_diff} (a)).
Even on the affected side, the acceleration amplitude of the asymmetric vibration stimulus was substantially higher than the measured vibration detection threshold, suggesting that the vibration itself was detectable.
Therefore, the marked reduction in directional discrimination performance on the affected side cannot be readily explained by failure to detect the vibration.

Overall, the pulling illusion should be understood not as a mere vibrotactile sensation, but as a phenomenon in which asymmetric vibration is perceived as directional pulling, that is, as illusory pulling perception.
The present findings suggest that the emergence of illusory pulling perception involves processing beyond basic vibration detection.
Neurological functions underlying motor-related signs may be associated with the emergence of illusory pulling perception.

\subsection{Implications for investigating the mechanisms of the pulling illusion}
This study reveals that the mechanisms underlying the pulling illusion cannot be fully explained by vibrotactile sensation at the fingertip alone; therefore, the pulling illusion should be examined across a broader range of processes, from basic vibration detection to subsequent perceptual processing.
In general, vibration applied to the fingertip is detected by cutaneous mechanoreceptors, conveyed through ascending somatosensory pathways to the thalamus, and subsequently processed in cortical somatosensory and sensorimotor regions \cite{somato_pathway}.
The transmission and processing of somatosensory information along this pathway may contribute to the process through which asymmetric vibration is perceived as directional pulling.
Participants with hemiplegia in this study had lesions involving the thalamus, internal capsule, parietal cortex, and frontal cortex (TABLE~\ref{tab:p}).
Lesions involving the thalamus or internal capsule may have affected the transmission of somatosensory signals \cite{somato_pathway}, whereas parietal lesions may have affected the perceptual processing of the transmitted information \cite{parietal_tactile}.
This interpretation is consistent with a previous study showing an association between the pulling illusion and delayed activity in the contralateral parietal cortex \cite{havas}.
The observations in PD participants with  tremor also raise the possibility that somatosensory processing through basal ganglia--thalamo--cortical circuits may contribute to the emergence of the pulling illusion \cite{PD_1}.
Therefore, this study indicates that the pulling illusion should be investigated not simply as a sensation accompanying asymmetric vibration, but as a perceptual phenomenon involving multiple stages of processing from peripheral input to the emergence of directional perception.

\subsection{Limitations and future work}
Several limitations of this study should be noted.
First, this study included participants with a wide range of neurological disorders, resulting in a limited sample size for each individual disorder.
While this heterogeneity constrains disorder-specific interpretations, it enabled the examination of illusion characteristics across diverse functional profiles.
The present findings may therefore help guide more targeted studies of conditions such as hemiplegia and PD with tremor in future work.
Second, motor-related signs were defined across participants to enable statistical analysis; however, this functional grouping may not capture more detailed, disorder-specific symptoms or individual variability.
In addition, factors that were not included as variables in the present model may also influence directional discrimination performance for the illusion.
Third, although vibration detection threshold was used as an index of peripheral vibrotactile sensitivity, the present study did not directly assess peripheral nerve function.
Therefore, the absence of a robust association between vibration threshold and directional discrimination performance should not be interpreted as evidence that peripheral sensory function is unrelated to the illusion.
Further investigation using more direct measures of peripheral function, such as nerve conduction studies, will be needed to clarify this relationship.
Finally, two participants were excluded from the clustering and regression analyses because they showed markedly below-chance performance on one tested side.
The origin of these response patterns remains unclear and may reflect procedural factors, a reversal of perceived pulling direction, or individual neurological characteristics.
Although these cases do not alter the central finding that directional discrimination performance cannot be fully explained by vibration detection threshold alone and is associated with motor-related signs, these exceptional cases should be further explored to better understand illusory pulling perception.

\section{Conclusion}
This study examined whether peripheral vibrotactile sensitivity alone is sufficient for the pulling illusion to emerge or whether functions associated with neurological disorders are also involved.
Directional discrimination performance was not consistently associated with vibration detection threshold, but was more consistently associated with motor-related signs.
Participants with hemiplegia also showed reduced performance on the affected side despite vibration amplitudes well above their detection thresholds.
These findings suggest that the pulling illusion is not merely a vibrotactile sensation, but an illusory pulling perception involving processing beyond basic vibration detection.

\end{document}